\documentclass[5p,authoryear]{elsarticle}

\usepackage{graphicx}
\usepackage{amssymb,amsthm}

\begin{document}

\begin{frontmatter}

\title{Spike timing-dependent plasticity induces non-trivial topology in the
brain}

\author{R. R. Borges$^{1,2}$, F. S. Borges$^1$, E. L. Lameu$^1$, A. M. 
Batista$^{1,3,4}$, K. C. Iarosz$^4$, I. L. Caldas$^4$, C. G. Antonopoulos$^5$,
M. S. Baptista$^6$}
\address{$^1$P\'os-Gradua\c c\~ao em Ci\^encias, Universidade Estadual de 
Ponta Grossa, Ponta Grossa, PR, Brazil.}
\address{$^2$Departamento de Matem\'atica, Universidade Tecnol\'ogica Federal do
Paran\'a, 86812-460, Apucarana, PR, Brazil.}
\address{$^3$Departamento de Matem\'atica e Estat\'istica, Universidade
Estadual de Ponta Grossa, Ponta Grossa, PR, Brazil.}
\address{$^4$Instituto de F\'isica, Universidade de S\~ao Paulo, S\~ao Paulo,
SP, Brazil.}
\address{$^5$Department of Mathematical Sciences, University of Essex, Wivenhoe
Park, UK.}
\address{$^6$Institute for Complex Systems and Mathematical Biology, Aberdeen,
SUPA, UK.}

\cortext[cor]{Corresponding author: antoniomarcosbatista@gmail.com}

\begin{abstract}
We study the capacity of Hodgkin-Huxley neuron in a network to change 
temporarily or permanently their connections and behavior, the so called spike
timing-dependent plasticity (STDP), as a function of their synchronous 
behavior. We consider STDP of excitatory and inhibitory synapses driven 
by Hebbian rules. We show that the final state of networks evolved by a STDP
depend on the initial network configuration. Specifically, an initial
all-to-all topology envolves to a complex topology. Moreover, external 
perturbations can induce co-existence of clusters, those whose neurons are
synchronous and those whose neurons are desynchronous. This work reveals that
STDP based on Hebbian rules leads to a change in the direction of the synapses
between high and low frequency neurons,  and therefore, Hebbian learning can
be explained in terms of preferential attachment between these two diverse
communities of neurons, those with low-frequency spiking neurons, and those
with higher-frequency spiking neurons.
\end{abstract}

\begin{keyword}
plasticity \sep synchronization \sep network
\PACS 87.10Hk \sep 87.19.lj \sep 87.19.lw
\end{keyword}

\end{frontmatter}

%%%%%%%%%%%%%%%%%%%%%%%%%%%%%%%%%
%%%%%%%%%%%%%%%%%%%%%%%%%%%%%%%%%

\section{Introduction}

Neuroplasticity, also known as brain plasticity or brain malleability
\citep{strong98,brenner00}, refers to the ability of the brain to reorganize 
neural pathways in response to new information, environment, development, 
sensory stimulation, or damage \citep{draganski04,james1890,lashley23}.
The term neuroplasticity was firstly introduced in 1948 by neuroscientist 
J. Konorski in a work \citep{konorski48} that showed the associative learning 
as a result of the adaptation of the brain to external stimuli. In 1949, D. O. 
Hebb, in his book entitled ``The Organization of Behavior'' \citep{hebb49}, 
proposed a plasticity rule, today known as Hebb's rule.

Scientific advances in neuroimaging and in noninvasive brain stimulation have 
provided insights to understand better neuroplasticity. Learning-induced 
structural alterations in gray and white matter have been documented in human 
brain \citep{dayan11}. Draganski and collaborators \citep{draganski04} used 
whole-brain magnetic-resonance imaging to observe learning-induced 
neuroplasticity. They verified structural changes in areas of the brain 
associated with the processing and storage of complex visual motion. Lu and 
collaborators \citep{lu03} demonstrated that neuroplasticity is affected by 
environmental stimuli. In addition, neuroimaging studies have showed 
alterations of neuroplasticity in depression, namely depressive disorder may be
associated with impairment of neuroplasticity \citep{fuchs04}.

Aiming at understanding the fundamental mechanisms behind plasticity, Popovych 
and collaborators studied the effect of noise on synchronous behavior in 
globally-coupled spiking Hodgkin-Huxley neurons with spike timing-depen\-dent
plasticity (STDP) and excitatory synapses \citep{popovych13,borges16}. STDP 
networks have nodes that adapt their synaptic strength according to some rule 
based on their spike timings \citep{gilson10,markram11,markram12}. Abarbanel
and Talathi \citep{abarbanel06} studied a neural circuit responsible for
recognizing interspike interval sequen\-ces by means of STDP of inhibitory
synapses. Similar results, though using different kinds of neural models, 
have been reported earlier by Kalitzin and collaborators \citep{kalitzin00}, 
where it was shown that coherent input can enhance synapses inducing high 
connectivity, whereas mutually anti-correlated inputs to individual neurons 
wea\-kens connectivity. On the contrary, the work in \citep{popovych13} however
shows that a fully uncorrelated input can enhance connectivity. Sadeh and
collaborators \citep{sadeh15} studied the emergence of functionally specific
connectivity in the visual cortex with Hebbian plasticity based on visual
experience. They showed that plasticity can lead to functionally specific and
stable connections in random networks composed of leaky integrate-and-fire
neurons. In our work, we focus the attention on the network dependence on
plasticity. To do that, we consider an initial all-to-all topology and focus
on the changes in synchronous and non synchronous states caused in a
Hodgkin-Huxley neural network with excitatory (eSTDP) and inhibitory synapses
(iSTDP).

Neural spike synchronization is responsible for information transfer 
\citep{chris15,baptista16}, and can be associated with forms of dysfunction. 
For instance, abnormally synchronized oscillatory activity has been reported
in Parkinson's disease \citep{hammond07}, epilepsy \citep{uhlhaas06} and some
other neurological disorders. Synchronous behavior was analised in systems of
synfire chains to solve binding problems. It was verified that dynamics of
binding may be modeled by competitive synchronization among synfire chains
\citep{abeles04,hayon05}. Moreover, synfire chains have been considered to
describe information transfer phenomena and coherent spiking \citep{wang16}. 

Our main goal is to show that spike timing-dependent plasticity of excitatory
and inhibi\-to\-ry synapses induces non-trivial topologies in the plastic
brain. Initial networks of neurons fully connected, evolve to a non trivial
complex network. Consequently, this non-trivial topology alters the synchronous
behavior. In our results, we have observed for some parameter conditions not
only the improvement of neural spiking synchronization, but also for other
parameter conditions that promote desynchronization. We have also observed
concurrent synchronous and non synchronous behavior in the neurons of a network
constructed for a particular set of parameters. Therefore, the onset of
synchronicity comes along side with desynchronicity in the plastic brain. This
balance between different synchronous behaviors is vital to maintain a
fundamental property of a brain network. Clusters need to be sufficiently
synchronous for information to be efficiently exchanged, but at the same time
sufficiently desynchronous to behave independently. Finally, we show that when
there is an external perturbation, the plastic neural network has an abrupt
change in behavior characterized by a first-order transition.

This paper is organized as follows: In Section II we introduce the neural 
network described by coupled Hodgkin-Huxley neural model. In Section III, we 
discuss our results about neural synchronization considering eSTDP and iSTDP.
In the last Section, we draw our conclusions.

\section{Neural network}

In this work,  we focus on eSTDP and iSTDP  based on Hebbian theory proposed in
Ref. \citep{hebb49}. These plastic mechanisms consist of synapses that become 
stron\-ger or weaker depending on the pre and postsynaptic neurons' activity. 
We have considered an initial network with a global coupling, with chemical
synapses where the connections are unidirectional, and the local dynamics is
described by the Hodgkin-Huxley model 
\citep{izhikevich2004,hodgkin1952}. The system is given by 
\begin{eqnarray}
C\dot{V_i} & = & I_i-g_{\rm K}n_i^{4}(V_i-E_{\rm K})-
g_{\rm Na}m_i^{3}h_i(V_i-E_{\rm Na}) \nonumber \\
& & -g_{L}(V_i-E_{\rm L})+\frac{(V_r^{\rm Exc}-V_i)}{\omega_{\rm Exc}}
\sum_{j=1}^{N_{\rm Exc}}\varepsilon_{ij}s_j \nonumber \\
& & +\frac{(V_r^{\rm Inhib}-V_i)}{\omega_{\rm Inhib}}
\sum_{j=1}^{N_{\rm Inhib}}\sigma_{ij}s_j+\Gamma_i,\\
\dot{n}_i & = & \alpha_{n_i}(V_i)(1-n_i)-\beta_{n_i}(V_i)n_i,\\
\dot{m}_i & = & \alpha_{m_i}(V_i)(1-m_i)-\beta_{m_i}(V_i)m_i,\\
\dot{h}_i & = & \alpha_{h_i}(V_i)(1-h_i)-\beta_{h_i}(V_i)h_i,
\end{eqnarray}
where $C$ is the membrane capacitance ($\mu$F/cm$^2$), $V_i$ is the membrane 
potential (mV) of neuron $i$ ($i=1,...,N$), $I_i$ is a constant current density
randomly distributed in the interval $[9.0,10.0]$, $\omega_{\rm Exc}$ 
(excitatory) and $\omega_{\rm Inhib}$ (inhibitory) are the average degree 
connectivities, $\varepsilon_{ij}$ and $\sigma_{ij}$ are the excitatory and 
inhibitory coupling strengths from the presynaptic neuron $j$ to the 
postsynaptic neuron $i$ \citep{gray}. The $\varepsilon_{ij}$ values are in the
interval $[0,0.5]$ and the $\sigma_{ij}$ values are in the interval
$[0,2\sigma_M]$. In our simulations, the maximum value for $\varepsilon_{ij}$ is
equal to 0.5 according to Reference \citep{popovych13}, and we consider the
maximum value for $\sigma_M$ equal to $0.75$ due to the fact that for
$\sigma_{ij}<1.5$ we observe a transition from synchronized to desynchronized
states. In addition, we have discarded a transient of $1.95\times 10^6$ms. We
consider that $80\%$ of the neurons are excitatorily coupled ($N_{\rm Exc}$) and
$20\%$ of them are inhibitorily coupled ($N_{\rm Inhib}$) according to anatomical
estimates for the neocortex \citep{noback05}. Both populations receive input
from all other neurons in own population and from the other population. We also
consider an external perturbation $\Gamma_i$, so that each neuron randomly
chosen receives an input with a constant intensity $\gamma=10\mu$A/cm$^2$
during $1$ms. In each time step $t_{\rm step}=0.01$ms a random input with
amplitude $\gamma$ is applied to each neuron with a probability equal to
$t_{\rm step}/14$, where $14$ms approximately corresponds to the inter-spike
interval of the Hodgkin-Huxley neuron. Functions $m(V_i)$ and $n(V_i)$
represent the activation for sodium and potassium, respectively, and $h(V_i)$
is the function for the inactivation of sodium. Functions
$\alpha_{n}$, $\beta_{n}$, $\alpha_{m}$, $\beta_{m}$, $\alpha_{h}$, $\beta_{n}$ are
given by
\begin{eqnarray}
\alpha_{n}(v) & = & \frac{0.01 v + 0.55}{1 - \exp \left(-0.1 v-5.5 \right)},\\
\beta_{n}(v) & = & 0.125\exp\left(\frac{-v-65}{80}\right),\\
\alpha_{m}(v) & = & \frac{0.1 v + 4}{1 - \exp\left (-0.1 v - 4\right)},\\
\beta_{m}(v) & = & 4\exp\left(\frac{-v-65}{18}\right),\\
\alpha_{h}(v) & = & 0.07\exp\left(\frac{-v-65}{20}\right),\\
\beta_{h}(v) & = & \frac{1}{1 + \exp\left(-0.1 v - 3.5\right)},
\end{eqnarray}
where $v=V/[mV]$. Parameter $g$ is the conductance and $E$ the reversal
potentials for each ion. Depending on the value of the external current density
$I_i$ ($\mu$A/cm$^2$) the neuron can present single spike activity or periodic
spikings. In the case of periodic spikes, if the constant $I_i$ increases, the
spiking frequency also increases. In this work, we consider
$C=1$ $\mu$F/cm$^{2}$, $E_{\rm Na}=50$mV, $E_{\rm K}=-77$ mV, $E_{\rm L}=-54.4$ mV,
$g_{\rm Na}=120$ mS/cm$^{2}$, $g_{\rm K}=36$ mS/cm$^{2}$, $g_{\rm L}=0.3$ mS/cm$^{2}$.
The neurons are excitatorily coupled with a reversal potential
$V^{\rm Exc}_{r}=20$mV, and inhibitorily coupled with a reversal potential
$V^{\rm Inhib}_{r}=-75$mV. The presynaptic potential $s_{i}$ is given by 
\citep{destexhe1994,golomb1993} 
\begin{equation}
\frac{ds_{i}}{dt}=\frac{5(1-s_{i})}{1+\exp\left(-\frac{v_{i}+3}{8}\right)}-
s_{i},
\end{equation}
where $v_i=V_i/[mV]$.

One of the key principles of behavioral neuroscience is that experience can
modify the brain structure, what is known as neuroplasticity \citep{ramon1928}.
Although the idea that experience may modify the brain structure can probably 
be traced back to the 1890s \citep{bliis1973,bliss1993}, it was Hebb who made 
this a central feature in his neuropsychological theory \citep{hebb1961}. With 
this in mind, we consider excitatory and inhibitory spike timing-dependent
plasticity according to the Hebbian rule. The coupling strengths 
$\varepsilon_{ij}$ and $\sigma_{ij}$ are adjusted based on the relative timing 
between the spikes of presynaptic and postsynaptic neurons 
\citep{bi1998,haas2006}. 

The plasticity dynamics can be mathematically defined as
\begin{equation}\label{edo}
{d\Delta\varepsilon(t) \over dt}=f(\Delta\varepsilon,V,t),
\end{equation}
where $\Delta\varepsilon$ is the update value of the synaptic weight.
Kalit\-zin and collaborators \citep{kalitzin00} considered a function $f$ that 
depends on the activation of the synapse, the transmembrane potential of the 
postsynaptic neuron, and the thresholds for switching on long-term 
potentiation and the long-term depression \citep{artola90}. In this work, we 
consider a linear function $f$ of the form 
$f(\Delta\varepsilon,t)=(a+c/t)\Delta\varepsilon$. The solution to 
the differential equation, Eq. (\ref{edo}), is given by 
$\Delta\varepsilon=bt^c\exp(at)$, where $a$, $b$ and $c$ are constants. For 
$c=0$ and $c\neq 0$, eSTDP and iSTDP are obtained, respectively. The plasticity
dynamics introduced by means of this linear function is not fundamentally 
related to physiological processes \citep{artola90}, but, by means of this 
function it is possible to find a fit that describes experimental results of 
eSTDP and iSTDP obtained in Refs. \citep{bi1998} and \citep{haas2006}.

\begin{figure}[htbp]
\begin{center}
\includegraphics[height=9cm,width=8cm]{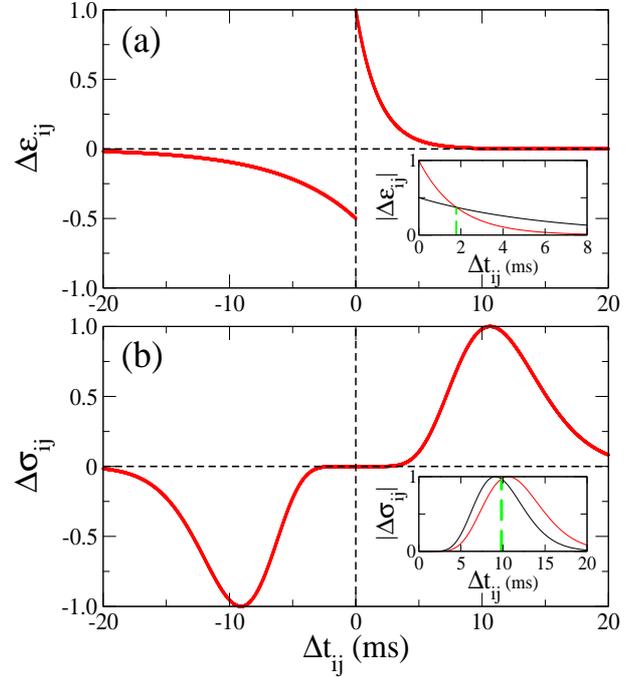}
\caption{(Color online) Plasticity as a function of the difference of spike
timing of post and presynaptic (a) excitatory (eSTDP) and (b) inhibitory
(iSTDP) synapse. The insets show the absolute value of the plasticity function.}
\label{fig1}
\end{center}
\end{figure}

The excitatory eSTDP is given by
\begin{equation}\label{eqplast}
\Delta \varepsilon_{ij}= \left\{
\begin{array}{ll}
\displaystyle A_{1}\exp(-\Delta t_{ij}/\tau_{1})\;,\;\Delta t_{ij}\geq 0 \\
\displaystyle -A_{2}\exp(\Delta t_{ij}/\tau_{2})\;,\;\Delta t_{ij} < 0 
\end{array}
\right. ,
\end{equation}
where $\Delta t_{ij}=t_{i}-t_{j}=t_{\rm pos}-t_{\rm pre}$, $t_{\rm pre}$ is the spike
time of the presynaptic and $t_{\rm pos}$ the spike time of the postsynaptic
neuron. Figure \ref{fig1}(a) exhibits the result obtained from 
Eq. (\ref{eqplast}) for $A_{1}=1.0$, $A_{2}=0.5$, $\tau_{1}=1.8$ms, and 
$\tau_{2}=6.0$ms. The initial synaptic weights $\varepsilon_{ij}$ are normally 
distributed with mean and standard deviation equal to $\varepsilon_M=0.25$ and 
0.02, respectively ($0\leq \varepsilon_{ij}\leq 2\varepsilon_M$). Then, they are
updated according to Eq. (\ref{eqplast}), where 
$\varepsilon_{ij}\rightarrow \varepsilon_{ij}+10^{-3}\Delta\varepsilon_{ij}$. The
insets in Fig. 1 show the absolute value of the plasticity function, where
the red and black lines are the potentiation and depression values,
respectively, as a function of $\Delta t_{ij}$. The green dashed line in the
inset Figures denotes the $\Delta t_{ij}$ value at wich the curves of
potentiation and depression intersect. The inset in Fig. 1(a) shows that for
$\vert \Delta t_{ij}\vert <1.8$ms the potentiation of $\varepsilon_{ij}$ is
bigger than the depression. Whereas in the case of iSTDP (inset in Fig. 1(b))
the potentiation of $\sigma_{ij}$ is bigger than the depression for
$\vert \Delta t_{ij}\vert >9.8$ms.

For the inhibitory iSTDP synapses, the coupling stren\-gth $\sigma_{ij}$ is 
adjusted based on the equation
\begin{equation}\label{eqplastI}
\Delta \sigma_{ij} =  \frac{g_0}{g_{\rm norm}} {\alpha}^{\beta} |\Delta t_{ij}| 
{\Delta t_{ij}}^{\beta -1} \exp(-\alpha |\Delta t_{ij}|),
\end{equation}
where $g_0$ is the scaling factor accounting for the amount of change in 
inhibitory conductance induced by the synaptic plasticity rule, and
$g_{\rm norm} = {\beta}^{\beta}  \exp(-\beta)$ is the normalizing constant.
Figure \ref{fig1}(b) exhibits the result obtained from Eq. (\ref{eqplastI}) for 
$g_0 = 0.02$, $\beta=10.0$, $\alpha =0.94$ if $\Delta t_{ij}>0$, and for
$\alpha=1.1$ if $\Delta t_{ij}<0$ \citep{talathi2008}. As a consequence,
$\Delta \sigma_{ij}>0$ for $\Delta t_{ij}>0$, and $\Delta \sigma_{ij}<0$ for
$\Delta t_{ij}<0$. The initial inhibitory synaptic weights $\sigma_{ij}$ are
normally distributed with mean and standard deviation equal to $\sigma_M$ and
0.02, respectively ($0\leq \sigma_{ij}\leq 2\sigma_M$). Then, the coupling
strengths are updated according to Eq. (\ref{eqplastI}), where
$\sigma_{ij}\rightarrow \sigma_{ij}+10^{-3}\Delta\sigma_{ij}$. The updates for
$\varepsilon_{ij}$ and $\sigma_{ij}$ are applied for the last postsynaptic spike.

\section{Spiking neuron synchronization}

To study the effect of plasticity on the neural network, we have calculated
the coupling strengths, and used the time-average order-parameter as a
probe of spike synchronization, a quantity expressed by 
\begin{equation}
R=\frac{1}{t_{\rm final}-{t_{\rm initial}}}\sum_{t_{\rm initial}}^{t_{\rm final}}
\left| \frac{1}{N}\sum_{j=1}^{N}\exp(i\psi_{j})\right| ,
\end{equation} 
where $t_{\rm final}-t_{\rm initial}$ is the time window for our estimation and the
phases are calculated by
\begin{equation}
\psi_{j}(t)=m+\frac{t-t_{j,m}}{t_{j,m+1}-t_{j,m}},
\end{equation}
where $t_{j,m}$ represents the time when a spike $m$ ($m=0,1,2,\dots$) in 
neuron $j$ occurs ($t_{j,m}< t < t_{j,m+1}$), with the beginning of each spike 
being when $V_j>0$. In synchronous behavior, the order-parameter magnitude 
approaches uni\-ty. In addition, if the spike times are uncorrelated, the 
order-parameter magnitude is typically small and vanishes for 
$N\rightarrow\infty$. When identical neurons are coupled, the neural network 
may exhibit complete  synchronization among spiking neurons, in other words, 
all other neurons may present identical time evolution of their action 
potentials. In this work, we are not considering identical neurons, and as 
result it is not possible to observe complete synchronization. However, an 
almost-complete synchronization may be observed.

\begin{figure}[htbp]
\begin{center}
\includegraphics[height=8cm,width=8cm]{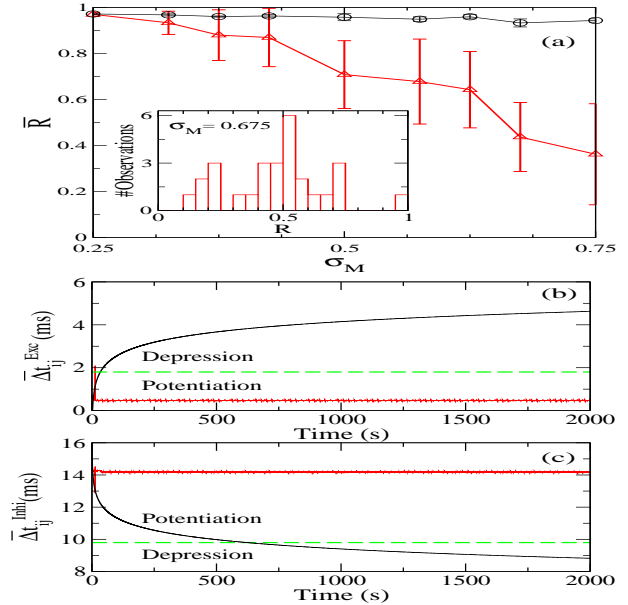}
\caption{(Color online) (a) Mean order-parameter ${\bar R}$ versus $\sigma_M$ 
for $\gamma=0.0$ and $\varepsilon_M=0.25$, a result without STDP (black 
circles) and the other one with STDP (red triangles). The bar is the standard 
deviation for $30$ different initial conditions. In the inset we consider 
$\sigma_M=0.675$. Panels (b) and (c) exhibit the time evolution of the 
average time-difference for excitatory and inhibitory connections, 
respectively. The black and red lines , for $\sigma_M=0.675$, correspond to 
${\bar R}\approx 0.1$ and ${\bar R}\approx 1$, respectively. The green line 
represents the separation between potentiation and depression.}
\label{fig2}
\end{center}
\end{figure}

Figure \ref{fig2}(a) shows the mean order-parameter ($\bar{R}$) that is 
calculated for different initial conditions, as a function of the inhibitory 
coupling strength $\sigma_M$ for a neural network with excitatory and 
inhibitory synapses, where we consider one case without STDP (black circles) 
and another with STDP (red triangles). For $\varepsilon_M$ equal to $0.25$ and 
varying $\sigma_M$, we do not observe a significant alteration of the $\bar{R}$
value  without STDP, due to the fact that initially the network has an 
all-to-all topology. Nevertheless, considering STDP we verify that the 
$\bar{R}$ values decrease with the increase of $\sigma_M$ and present a  
large standard deviation. This standard deviation occurs due to the existence 
of different synchronization states. Then, both the upper border of the
inhibitory coupling $2\sigma$ and the different initial conditions are
important to change the dynamics of the network with STDP and without external
perturbation. This is verified by means of the decay of the $R$ values and the
large standard deviation bar. In the inset (Fig. \ref{fig2}(a)), we consider
$\sigma_M=0.675$ and calculate the order-parameter for different initial
conditions. As a result, we can see a distribution presenting different
synchronization states, including desynchronization and synchronization. In 
Figs. \ref{fig2}(b) and \ref{fig2}(c) we consider $\sigma_M=0.675$ according 
to the inset, and calculate the time evolution of the average time-difference 
for excitatory and inhibitory connections,
\begin{eqnarray}
\bar{\Delta t}_{ij}^{\rm Exc} & = & \frac{1}{\tau}\sum_{i\neq j}
| t_{\rm pre}^{\rm Exc}-t_{\rm pos} |, \\
\bar{\Delta t}_{ij}^{\rm Inhib} & = & \frac{1}{\tau}\sum_{i\neq j}
| t_{\rm pre}^{\rm Inhib}-t_{\rm pos} |, 
\end{eqnarray}
respectively, for different configurations of the initial networks and
$\tau=100$ms. The black line shows the case in which the network goes to a
desynchronized state (${\bar R}\approx 0.1$), whereas the red line exhibits
the case of a network that presents synchronous behavior (${\bar R}\approx 1$).
In both cases, we consider the same parameters, except the seed to generate the
random distribution of the constant current density $I_i$. Through Figures
\ref{fig2}(b) and \ref{fig2}(c)  it is possible to verify why and when the
coupling matrix suffer substantial changes. The transition occurs when the 
black or red curves cross the green line. At this time, depreciation induces
weak strength in the coupling matrix, and potentiation induces strong 
strength.

\begin{figure}[htbp]
\begin{center}
\includegraphics[height=7cm,width=7cm]{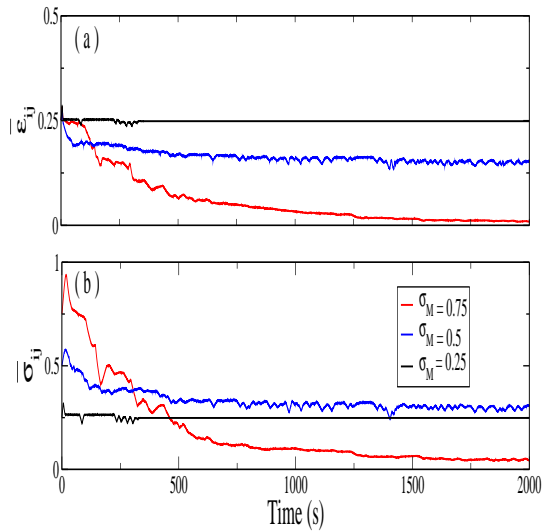}
\caption{(Color online) Time courses of the mean (a) excitatory and (b)
inhibitory coupling strengths from some regimes showed in Figure \ref{fig2}(a).}
\label{fig3}
\end{center}
\end{figure}

\begin{figure}[htbp]
\begin{center}
\includegraphics[height=8cm,width=8cm]{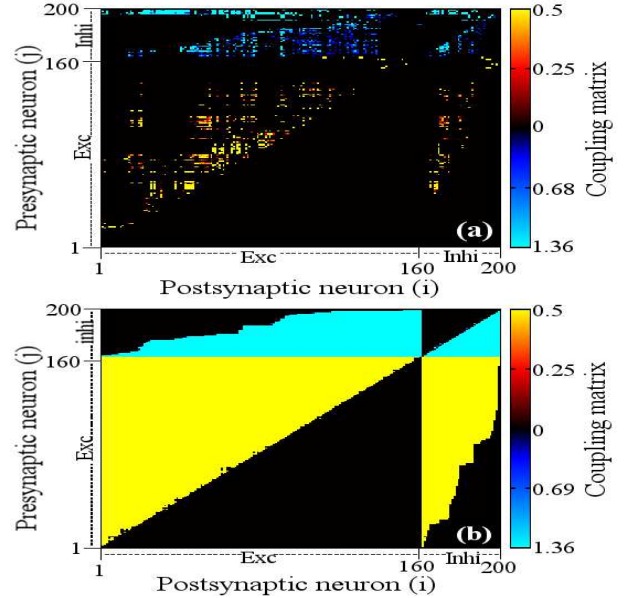}
\caption{(Color online) Coupling matrix for $\gamma=0.0$, $\varepsilon_M=0.25$, 
and $\sigma_M=0.675$, where we choose values for parameters to provide
the cases for (a) ${\bar R}\approx 0.1$ showing many uncoupled neurons, and (b) 
${\bar R}\approx 1$ exhibiting many directed couplings, according to the inset 
in Fig. \ref{fig2}(a). The synaptic weights are encoded in color, where the 
maximum value for $\varepsilon_{ij}$ (yellow) is $0.5$ and the maximum value 
for $\sigma_{ij}$ (blue) is $1.36$. The neurons are ordered according to
$I_i\leq I_j$ for $i<j$.}
\label{fig4}
\end{center}
\end{figure}

Figure \ref{fig3} exhibits the time courses of the mean excitatory (Fig.
\ref{fig3}(a)) and inhibitory (Fig. \ref{fig3}(b)) coupling strengths from the
multiple coexisting regimes that are shown in Figure \ref{fig2}(a). We see 
that for $\sigma_M=0.25$ both $\bar{\varepsilon}_{ij}$ and $\bar{\sigma}_{ij}$ 
have constant values for the time approximately greater than $700$s, and the 
learning produces a triangular-type connecting matrix (as shown in Fig.
\ref{fig4}), meaning that the connections among all neurons become
preferentially directed. For $\sigma_M=0.5$ the $\bar{\varepsilon}_{ij}$ values
decrease to approximately $0.15$, while $\bar{\sigma}_{ij}$ values oscillate
about $0.25$, and the coupling matrix becomes partitioned, indicating the
existence of larger clusters. Increasing the upper border $\sigma_M$ to $0.75$
both $\bar{\varepsilon}_{ij}$ and $\bar{\sigma}_{ij}$ tend to $0$, and the
coupling matrix becomes sparse.

In Fig. \ref{fig4}, the synaptic weights $\varepsilon_{ij}$ and $\sigma_{ij}$ 
are encoded in color for $\gamma=0.0$, $\varepsilon_M=0.25$, and 
$\sigma_M=0.675$, where we choose values of the parameters that
provide the cases for (a) ${\bar R}\approx 0.1$ and (b) ${\bar R}\approx 1$ 
according to the inset in Fig. \ref{fig2}(a). The synaptic weights are 
suppressed in the desynchronized regime (Fig. \ref{fig4}(a)), and consequently 
the coupling matrix presents a small number of connections. This behavior 
can be verified by means of the black lines in Figs. \ref{fig2}(b) and 
\ref{fig2}(c). In addition, the synaptic weights are potentiated (red lines in 
Figs. \ref{fig2}(b) and \ref{fig2}(c)) in the synchronized regime (Fig. 
\ref{fig4}(b)), and the coupling matrix exhibits a triangular shape. We have 
verified that, in this case, the synchronous behavior has a dependence on the 
direction of synapses. In other words, when the presynaptic neurons are
excitatory the synapses from the high frequency to the low frequency neurons
become stronger. When the presynaptic neurons are inhibitory, the synapses
from the low frequency to the high frequency neurons become stronger.

Figure 4 shows the final topologies for two networks initially set with a
global coupling topologies after being evolved by a STDP proccess. We see that
the STDP induces a non-trivial topology in the network resulting in networks
sparsely connected, moderately connected (Fig. 4(a)), or densely connected with
strong preferential attachment (Fig. 4(b)).

\begin{figure}[htbp]
\begin{center}
\includegraphics[height=8cm,width=8cm]{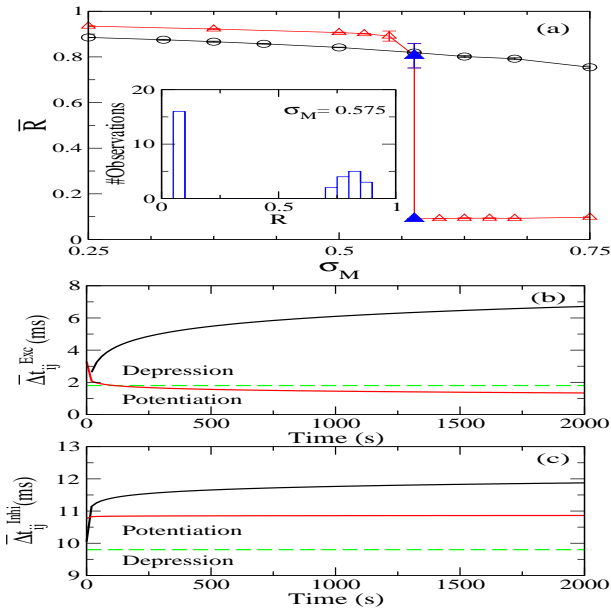}
\caption{(Color online) (a) Mean order-parameter versus $\sigma_M$ for 
$\gamma=10.0$ $\mu\rm A/cm^2$, $\varepsilon_M=0.25$, a result without 
STDP (black circles) and another one with STDP (red triangles). Inset plot for 
$\sigma_M=0.575$ (blue triangles) and $30$ values of $R$, where each $R$ is
calculated from a different initial configuration. Figures (b) and (c) exhibit
the time evolution of the average time-difference for excitatory and inhibitory
connections, respectively, where $\sigma_M$ is equal to $0.575$. The black and
red lines correspond to ${\bar R}\approx 0.1$ and ${\bar R}\approx 0.8$,
respectively. The green dash represents the separation between potentiation and
depression.}
\label{fig5}
\end{center}
\end{figure}

Considering an external perturbation ($\Gamma_i>0$), we also study the cases 
without and with plasticity. In the case without STDP, we verify that the mean 
order-parameter has a small decay when $\sigma_M$ increases, as shown in Fig.
\ref{fig5}(a) with black circles. The red triangles represent the case with
STDP, and unlike the case without perturbation (Fig. \ref{fig2}(a)), there is
an abrupt transition (blue triangles), due to a first-order transition in the
average order parameter.  First-order transition is a term that comes from
Thermodynamics and here represents a discontinuity of the mean order-parameter
function with respect to the inhibitory coupling strength. In this case, the
upper border of the inhibitory coupling is relevant to produce alteration in
the dynamics, while the different initial conditions are important only at the
transition. Based on the results in the inset (Fig. \ref{fig5}(a)), we verify
that the network in the transtition can be either in one of the states: (i)
high ${\bar R}$ with potentiation of the average-time difference for excitatory
and inhibitory connections (red lines in Figs. \ref{fig5}(b) and
\ref{fig5}(c)), or (ii) low ${\bar R}$ with excitatory average time-difference
in the depression region and inhibitory in the potentiation region (black
lines).

The transition from the synchronized to the desynchronized states was reported
in studies on how stimulation impact on neurological disorders induced by an
abnormal neuronal synchronization \citep{tass06,popovych12}. A first order
transition was also observed in \citep{popovych13} when the stimulation
intensity varies in a neural network with eSTDP. In our simulations, we observe
the transition to desynchronization caused by a variation in the inhibitory
coupling in neural networks with both eSTDP and iSTDP.

\begin{figure}[htbp]
\begin{center}
\includegraphics[height=8cm,width=8cm]{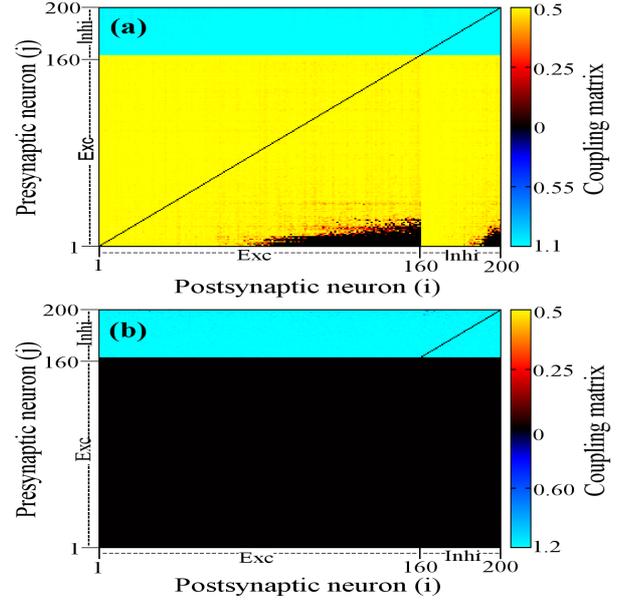}
\caption{(Color online) Coupling matrix for $\gamma=10.0$ 
$\mu\rm A/cm^2$, $\varepsilon_M=0.25$. (a) $\sigma_M=0.55$ 
(${\bar R}\approx 1$) showing a large quantity of coupled neurons, and (b) 
$\sigma_M=0.6$ (${\bar R}\approx 0.1$) exhibiting connections from
inhibithory to excitatory neurons. The synaptic weights are encoded in 
color, where the maximum value for $\varepsilon_{ij}$ (yellow) is $0.5$, 
the maximum value for $\sigma_{ij}$ (blue) is $1.1$ in (a) and $1.2$ in (b).
The neurons are ordered according to $I_i\leq I_j$ for $i<j$.}
\label{fig6}
\end{center}
\end{figure}

Figure \ref{fig6} illustrates the coupling matrix for the two states of the 
first-order transition. In Fig. \ref{fig6}(a), we can see the coupling 
configuration that corresponds to high ${\bar R}$. The network presents high 
connectivity, and for this reason it is possible to observe synchronous 
behavior. For the case of low ${\bar R}$, we verify that the network has only
connections from neurons belonging to the inhibitory population to any other
neuron, as shown in Fig. \ref{fig6}(b).

\section{Conclusion}
 
In conclusion, we have studied the effects of spike tim\-ing-dependent 
plasticity on the synchronous behavior and the evolved connecting topology of
neural networks constructed with Hodgkin-Huxley neurons. In our simulations, we
considered parameter values for the Hodgkin-Huxley system and STDP according to
experimental values found in the neuroscience literature
\citep{bi1998,haas2006}. Regarding the evolved topology, our main conclusion
is that learning under a STDP results in evolved networks that present complex
topology. Concerning the dynamic synchronous behavior of the evolved networks,
we observe that the studied networks exhibit concurrent synchronous and non
synchronous states with characteristics that depend on both the upper border of
the inhibitory coupling and the initial conditions. Specifically, we verify
that the main role of the inhibitory connections is to produce a delay in the
spiking time of the postsynaptic neurons. As a consequence, the increase of the
inhibitory coupling strength can suppress synchronous behavior, which
contributes to a decrease in the mean order parameter. Moreover, the transition
from low to a high synchronous state is smooth by alterations of the inhibitory
synapses. When a random external perturbation is introduced in the network,
this transition becomes discontinuous, i.e., we observe a first-order
transition. Similarly to the non-perturbed network, we also find coexistence of
synchronous and non-synchronous neurons in the perturbed networks. 

In future works, we plan to study synchronous states in the brain considering 
plasticity dynamics in terms of the thresholds for switching on the long-term 
potentiation and the long-term depression.  We also plan to investigate how the
final behavior of the network depends on the initial population of excitatory
neuron.

%%%%%%%%%%%%%%%%%%%%%%%%%%%%%%%%%%%%
%%%%%%%%%%%%%%%%%%%%%%%%%%%%%%%%%%%%

\section*{Acknowledgments}
This study was possible by partial financial support from the following 
Brazilian government agencies: CNPq, FAPESP (2011/19296-1, 2015/07311-7,  
2016/16148-5) and CAPES. AMB, KCI, CGA, and MSB partial support from
EPSRC-EP/I032606. We also wish thank Newton Fund and COFAP.

\bibliographystyle{elsarticle-harv}

\end{document}